\documentclass[twocolumn,english,aps,prd,reprint,floatfix,notitlepage,nofootinbib,preprintnumbers,superscriptaddress]{revtex4-1}
\pdfoutput=1
\usepackage{lmodern}

\usepackage[T1]{fontenc}
\usepackage[latin9]{inputenc}
\usepackage{geometry}
\usepackage{subfigure,lmodern, amsmath,amssymb, graphicx, pifont, adjustbox, bm, xcolor}
\usepackage{amsfonts}
\usepackage{geometry}
\usepackage{comment}
\usepackage{mathtools}
\usepackage{float}
\usepackage{slashed}
\usepackage{ragged2e}
\usepackage{tikz}
\usepackage{array}
\usepackage{hhline}
\usepackage{dsfont}
\usepackage{braket}

\makeatletter\g@addto@macro\bfseries{\boldmath}\makeatother

\usepackage{stackengine}
\usepackage{esint}
\usepackage[unicode=true,pdfusetitle,
 bookmarks=true,bookmarksnumbered=false,bookmarksopen=false,
 breaklinks=false,pdfborder={0 0 1},backref=false,colorlinks=true]
 {hyperref}
\hypersetup{
 pdfauthor={Ivano Basile, Grant N. Remmen, Georgina Staudt},
 citecolor=black,linkcolor=black,urlcolor=black}

\newcommand{\appendixref}[1]{\hyperref[#1]{appendix~\ref{#1}}}
\def\equationautorefname~#1\null{eq.\,(#1)\null}
\usepackage{breakurl}
\usepackage{breakurl}
\usepackage[hang,flushmargin]{footmisc} 
\allowdisplaybreaks
\makeatletter

\usepackage{etoolbox}
\apptocmd{\thebibliography}{\justifying\setlength{\leftskip}{7.4mm}}{}{} 
 
 \usepackage{relsize}
\usepackage{babel}

\makeatletter
\def\simgt{\mathrel{\lower2.5pt\vbox{\lineskip=0pt\baselineskip=0pt
           \hbox{$>$}\hbox{$\sim$}}}}
\def\simlt{\mathrel{\lower2.5pt\vbox{\lineskip=0pt\baselineskip=0pt
           \hbox{$<$}\hbox{$\sim$}}}}
\makeatother

\usepackage{changepage}

\makeatletter
\def\l@subsubsection#1#2{}
\makeatother

\newcommand{\be}{\begin{equation}}
\newcommand{\ee}{\end{equation}}
\newcommand{\bea}{\begin{eqnarray}}
\newcommand{\eea}{\end{eqnarray}}

\newcolumntype{P}[1]{>{\centering\arraybackslash}p{#1}}

\definecolor{dartmouthgreen}{rgb}{0.05, 0.5, 0.06}

\usepackage[backgroundcolor=white!95!red, linecolor=cyan, bordercolor=cyan, textsize=footnotesize]{todonotes}

\newenvironment{eqaed}
    {\begin{equation}
    \begin{aligned}
    }
    { 
    \end{aligned}
    \end{equation}
    \ignorespacesafterend
    }

\begin{document}

\preprint{MPP-2026-24}

\title{Higher-Spin and Higher-Point Constraints on Stringy Amplitudes}

\author{Ivano Basile}
\affiliation{Max-Planck-Institut f\"ur Physik (Werner-Heisenberg-Institut), Boltzmannstra{\ss}e 8, 85748 Garching, Germany}
\author{Grant N.~Remmen}
\affiliation{Center for Cosmology and Particle Physics, Department of Physics, New York University, New York, NY 10003}
\author{Georgina Staudt}
\affiliation{Max-Planck-Institut f\"ur Physik (Werner-Heisenberg-Institut), Boltzmannstra{\ss}e 8, 85748 Garching, Germany}

\begin{abstract}

\noindent We employ multiparticle factorization to constrain deformations of tree-level open string amplitudes. Assuming minimal degeneracy among intermediate states of the same spin up through the second excited level, we find that the Regge intercept among all amplitudes of the Koba-Nielsen type can be uniquely fixed using seven-point factorization, precisely matching the bosonic string. Moreover, we produce novel constraints on deformations of the worldsheet integrand. We then turn to deformations of superstrings, with massless external states and arbitrary spectral degeneracy, using soft kinematics. Accounting for the infinite tower of higher-spin resonances, we obtain novel multipositivity bounds to leading and subleading order in the large-level limit. We apply these bounds to the simplest factorizable satellite deformation in the family of amplitudes found by Gross, showing that any deformation of four-point string amplitudes of this type is forbidden by unitarity. Our results reinforce the folklore that the higher-spin tower of string excitations is dramatically more rigid than any finite number of species.

\noindent 
\end{abstract}
\maketitle 

\section{Introduction}\label{sec:introduction}

The celebrated Veneziano amplitude \cite{Veneziano:1968yb} describes tree-level scattering of open strings in a way that satisfies causality, unitarity, and consistent factorization into three-point vertices. These features, describing an infinite tower of higher-spin states while retaining consistency and good behavior in the ultraviolet, are striking enough to suggest that the Veneziano amplitude might be somehow extremely special. Indeed, if string theory is {\it the} unique theory of quantum gravity, one would hope for a certain rigidness of the structure of its S-matrix. While in the full theory, the worldsheet CFT of course completely fixes the tree-level amplitudes in weak coupling limits, one could hope for signs of this uniqueness to be apparent from self-consistency alone.

However, unitarity and factorization alone are not enough. 
In recent years, a litany of other amplitudes have been found that represent apparently consistent deformations of string amplitudes in a way that honors consistency criteria like positivity of partial waves~\cite{Cheung_bootstrap, Cheung_dual_resonance, Bhardwaj:2024klc,Cheung_veneziano_variations, Maldacena_2022, Geiser_genveneziano, H_ring_superpolysoftness, Huang_triple_product,Cheung:2024uhn,Cheung:2024obl,Gross:1969db}. Similarly, bootstrap constraints seem to require high-energy input, e.g., extremely good UV behavior~\cite{cheung2025strings}---or alternatively information about the whole higher-spin tower~\cite{Albert:2024yap,Cheung:2024uhn,Cheung:2024obl}\footnote{See also Refs.~\cite{Guerrieri:2021ivu, Guerrieri:2022sod} for nonperturbative bootstrap constraints along similar lines.} or black-hole physics~\cite{Basile:2023blg, Bedroya:2024ubj, Herraez:2024kux}---in order to uniquely fix the answer.\footnote{See Refs.~\cite{Markou:2023ffh, Basile:2024uxn, Markou:2025xpf} for recent developments on the structure of the higher-spin string spectrum.} A well-known one-parameter $q$-deformation interpolating between the Veneziano amplitude and scalar field theory is the Coon amplitude \cite{coon_1969}, which has also been revisited and tested for viability in recent literature~\cite{Jepsen_2023, bhardwaj2023dualresonantamplitudesdrinfeld, geiser2024bakercoonromansnpointamplitudeexact, bhardwaj2024unitaritycoonamplitude}.
Recent analyses have shown, however, that going to higher-point and demanding consistent factorization leads to strong constraints on deformations of the string~\cite{Arkani-Hamed:2023jwn,Cheung_2025}. Indeed, recent bootstrap approaches have suggested that the string may indeed be unique once additional criteria beyond unitarity are implemented, e.g. SUSY or string monodromy~\cite{Berman:2023jys,Berman:2024eid,Berman:2024wyt,Elvang:2026pmc}. Indeed, it has recently been analytically shown~\cite{cheung2025strings}  that simply demanding the simplest amplitude consistent with ultrasoft UV scaling results in both the string spectrum and dynamics as outputs of a bootstrap.
In light of these results, it is worth returning to the problem of factorization and ask whether, among deformations of the string, we can uniquely pick out the Veneziano amplitude by imposing modest constraints on the degeneracy of states of its spectrum.

In order to specify the kinds of deformations we will study, we take a look at the mathematical formulation of the Veneziano amplitude. We start from the Koba-Nielsen integral~\cite{Koba:1969rw}
\begin{equation}
    A_\text{KN}(p_i) = \int \frac{d^N z}{SL(2, \mathbb{R})} \frac{\Pi_{i<j} z_{j,i}^{-2 p_i p_j}}{z_{1,2}...z_{N,1}},\label{eq:AKN}
\end{equation}
where $z_i \in (0,1) $ are the worldsheet moduli integrated over the disc, and $z_{i,j} = z_i - z_j$. Here, $N$ is the number of external particles, and the spectrum is defined by $m^2_n = n$. The amplitude is meromorphic, with poles in the planar Mandelstam invariants $X_{i,j} = (p_i + p_{i+1} +...+ p_{j-1})^2$ when an intermediate state goes on-shell. In order to make this dependence manifest, we introduce the cross-ratio $u$-variables~\cite{Gross:1969db,Koba:1969rw,BardakciRuegg,ChanTsou} $u_{i,j} = \frac{z_{i-1,j}z_{i,j-1}}{z_{i,j}z_{i-1,j-1}}$, from which we obtain
\begin{equation}
    A_\text{KN}(X_{i,j}) = \int \frac{d^N z}{SL(2, \mathbb{R})} \frac{\Pi_{i<j} u_{i,j}^{X_{i,j} + \alpha_0}}{z_{1,2}...z_{N,1}}.\label{eq:AKNu}
\end{equation}
Note that, here, we introduced a Regge intercept $\alpha_0$, so that the poles are at $X_{i,j} = -m_n^2$ for the shifted spectrum $m^2_n = n + \alpha_0$. For the bosonic string, the amplitudes are precisely as given in Eq.~\eqref{eq:AKNu} with $\alpha_0 =-1$, while for the open superstring, $\alpha_0=0$ and the amplitudes are dressed with a prefactor containing polarization data.
In order to expeditiously compute residues of the string and its deformations, we redefine the $u$-variables in terms of the positive $y$-variables of Refs.~\cite{Arkani-Hamed:2024nzc,BinGeom,Arkani-Hamed:2023lbd,Arkani-Hamed:2019mrd,Gluons,Zeros,Splits}, reviewed in Refs.~\cite{Arkani-Hamed:2023jwn,Cheung_2025}, in terms of which
\begin{equation}
    A_\text{KN}(X_{i,j}) = \int_0^\infty \prod_{{\cal C}\in {\cal T}}\frac{dy_{\cal C}}{y_{\cal C}} \prod_{i<j} u_{i,j}^{X_{i,j} + \alpha_0},\label{eq:AKNy}
\end{equation}
where the product is taken over all chords ${\cal C}$ in a triangulation ${\cal T}$ defining a factorization channel.
In terms of these new variables, the extraction of residues, which is a challenge in the traditional form of the Koba-Nielsen integral~\eqref{eq:AKN}, becomes a straightforward matter of computing poles when certain $y$-variables go to zero.

At this point, we can ask two questions. First, what are the allowed values of the Regge intercept? Partial wave unitarity of the four-point amplitude has been shown to allow a range of $\alpha_0$, also depending on the spacetime dimension $D$~\cite{Arkani-Hamed:2023jwn}.
If we make assumptions about the degeneracy of the spectrum, can we derive additional constraints by demanding factorization on excited states?

Second, we can introduce deformations to the Koba-Nielsen factor and see whether they can still be viable amplitudes, i.e., whether they obey causality, unitarity, and consistent factorization? This possibility will be implemented by 
\begin{equation}
    A_\text{KN}(X_{i,j}) = \int \frac{d^N z}{SL(2, \mathbb{R})} \frac{\Pi_{i<j} u_{i,j}^{X_{i,j} + \alpha_0}}{z_{1,2}...z_{N,1}} P_N (u),\label{eq:AKNPu}
\end{equation}
where the worldsheet deformation factor $P_N (u)$ is some function of the $u$-variables that obeys the requisite crossing symmetry and analyticity conditions.

In Ref.~\cite{Arkani-Hamed:2023jwn}, it was proposed to leverage factorization of deformed string amplitudes in order to probe their consistency. Any $N$-point string scattering process at levels $n_N$ for each intermediate momentum has to be factorizable consistently down to three-point processes with real couplings. Therefore, any deformation of the Veneziano amplitude failing to satisfy this condition is ruled out by unitarity. In Ref.~\cite{Arkani-Hamed:2023jwn}, this approach was tested for both massive and massless scattering with arbitrary Regge intercept through the first excited level ($n=1$), from which some particular families of string deformations were excluded. In this work, we extend these results by performing the factorization condition analysis for massive scattering up through the second excited level $(n=2)$, which---in the case of minimal degeneracy among states of the same spin---will force us to fix the Regge intercept to its known value of the bosonic string $\alpha_0 =-1$ and exclude a broader class of deformations than those probed in Ref.~\cite{Arkani-Hamed:2023jwn}. In the case where the lowest-lying states are massless (the superstring), however, one already encounters degeneracies in the spectrum at level $n=2$. In order to treat this case properly, one would need to account for all degeneracies already in the ansatz for the residues. While not impossible, this approach would lead to extremely complex algebraic systems involving an arbitrary number of couplings; such an analysis is beyond the scope of the present work. Instead, to treat deformations of the superstring, we take a different approach for massless scattering, namely constraining them via multipositivity bounds on residues recently discovered in Ref.~\cite{Cheung_2025}. These bounds on residues of scattering amplitudes with different numbers of external particles prove to be sufficiently powerful enough to not only impose positivity on the deformations, but also to rule out the four-point deformations originally proposed by Gross in Ref.~\cite{Gross:1987kza}, in a manner completely independent of the degeneracy of states.

The remainder of this paper is structured as follows. In Sec.~\ref{sec:level-two_constraints} we present our analysis of level-two factorization constraints for minimal degeneracy. We begin by only varying the Regge intercept $\alpha_0$ in Sec.~\ref{sec:level-two_regge}, showing that it must be equal to $-1$ as in the bosonic string. We then allow more general deformation of the worldsheet integrand and derive constraints associated with minimal degeneracy through mass level $n=2$ in Sec.~\ref{sec:level-two_Pu}. In order to constrain degenerate spectra, in Sec.~\ref{sec:massless_scattering} we fix $\alpha_0 = 0$ and apply the multipositivity bounds of Ref.~\cite{Cheung_2025} to the whole higher-spin tower of exchanges with a saddle-point analysis. In Sec.~\ref{sec:satellite} we apply the resulting bounds to the simplest deformation introduced in Ref.~\cite{Gross:1969db}, finding that {\it any} such four-point deformation of the string is excluded by unitarity.
We discuss future directions in Sec.~\ref{sec:conclusions}.

\section{Factorization at level two}\label{sec:level-two_constraints}

\subsection{Fixing the Regge intercept for minimal degeneracy}\label{sec:level-two_regge}

Let us first consider the planar amplitudes of open string theory, with arbitrary Regge intercept $\alpha_0$.
The exchanged states are at masses $m_n^2 = n+\alpha_0$, where the spin spectrum at level $n$ runs from $0$ to $n$, and the external particles are colored scalars of mass $\alpha_0$.
For $\alpha_0  = -1$, these amplitudes describe the scattering of $N$ tachyons in the bosonic string, while for $\alpha_0 = 0$, these amplitudes describe the scattering of massless scalars in the so-called $Z$-theory~\cite{Ztheory}, or, when dressed with external polarization data, the scattering of gluons in superstring theory.
The $N$-point string amplitude is given in the form of a Koba-Nielsen integral in Eq.~\eqref{eq:AKNu}, and we extract the residues using the positive $y$-variables discussed in Eq.~\eqref{eq:AKNy}.

We can use these residues to construct the fundamental three-point couplings of the string. This was accomplished up through level $n=1$ in Ref.~\cite{Arkani-Hamed:2023jwn} in the case of minimal degeneracy: at most one state at each given spin $\leq n$ for each mass level $n$.
There, it was found that unitarity---i.e., a consistent factorization into three-point amplitudes with real couplings---requires $\alpha_0 \geq -1$.
Notably, however, the bosonic string exhibits minimal degeneracy through level $n=2$ (see, e.g., Refs.~\cite{Markou:2023ffh,Pesando:2024lqa}).
Solving the analogous factorization problem through the second excited level, while imposing minimal degeneracy, we will find interesting additional constraints.
\begin{figure}[H]
    \centering
    \includegraphics[width=0.4\textwidth]{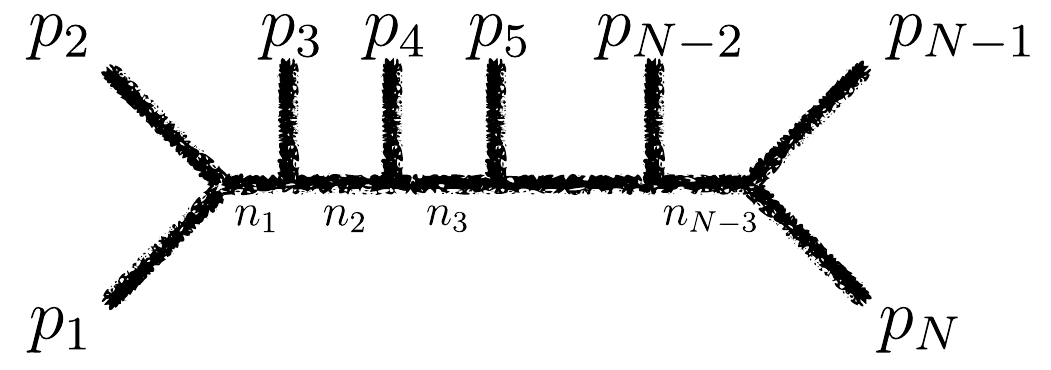}
    \caption{Conventions for the half-ladder topology of the factorization channels considered in this paper. Here, the Mandelstam $X_{1,i+2}$ is on shell at mass level $n_i$ for $i=1,\ldots,N-3$.}
    \label{fig:halfladder}
\end{figure}
We define the three-point amplitudes following the conventions of Ref.~\cite{Arkani-Hamed:2023jwn}. The residues for the half-ladder topologies depicted in Fig.~\ref{fig:halfladder} depend on the three-point amplitudes coupling a massless scalar to states of levels $n_{1}$ and $n_2$ of spin $s_1$ and $s_2$, respectively, for which there are $\min(s_1,s_2)$ independent couplings,
\begin{equation}\hspace{-2mm}
    \begin{aligned}
        &A_{n_1,n_2}^{s_1,s_2}(q_1,q_2) = \\ &\sum_{j=0}^{\min(s_1,s_2)}\!\!\lambda_{n_1,n_2}^{s_1,s_2;j}(-iq_{2} e_1)^{s_1-j}(iq_{1} e_2)^{s_2-j}(e_1 e_2)^j.
    \end{aligned}\hspace{-1mm}
\end{equation}

Here, we have denoted all Lorentz contractions among vectors within parentheses and have written the polarization tensors for states 1 and 2 as outer products of vectors $e_1^\mu$ and $e_2^\mu$, labeling the corresponding momenta by $q_1$ and $q_2$, with $q_1$ defined as incoming and $q_2$ as outgoing.\footnote{If we were to go beyond minimal degeneracy at each mass and spin, we would have to introduce additional flavor labels on $\lambda$ in the form $\lambda_{n_1,n_2}^{s_1,s_2;j}(f_1,f_2)$, where the ranges of the degeneracy labels $f_{1,2}$ depend on the mass and spin labels $n_{1,2}$ and $s_{1,2}$. The matching procedure works in precisely the same way, albeit much more laboriously.}
With these definitions, and recalling the $(-1)^{s_1+s_2}$ factor associated with swapping the color ordering, the coupling constants $\lambda_{n_1,n_2}^{s_1,s_2;j} = \lambda_{n_2,n_1}^{s_2,s_1;j}$ are symmetric on $1\leftrightarrow 2$ and, crucially, are {\it real} by virtue of unitarity.
Building all four-, five-, and six-point half-ladder topologies for all combinations of legs at levels $n \in \{0,1,2\}$, we find the ansatz residues in terms of the $\lambda$ couplings, the Regge intercept $\alpha_0$ and the spacetime dimension $D$. The four-point expressions are
\begin{equation}
\begin{aligned}
    & R_0^\lambda = (\lambda_{0,0}^{0,0;0})^2 \, , \\
    & R_1^\lambda = (\lambda_{0,1}^{0,0;0})^2 - \frac{(\lambda_{1,1}^{0,0;0})^2(2X_{2,4}+3\alpha_0-1)}{4} \, , \\
    & R_2^\lambda = (\lambda_{0,2}^{0,0;0})^2 - \frac{(\lambda_{1,2}^{0,0;0})^2(2X_{2,4}+3\alpha_0-2)}{4} \\
    &\qquad + (\lambda_{2,2}^{0,0;0})^2 \, \frac{(D-2)(3\alpha_0-2)^2}{16(D-1)} \\
    & \qquad + \frac{(\lambda_{2,2}^{0,0;0})^2}{4} \, X_{2,4}(3\alpha_0-2+X_{2,4}) .
\end{aligned}\label{eq:ansatz_residues_4pt}
\end{equation}
The higher-point expressions are much more involved, and are given in the supplemental files \cite{supplementRansatz} through seven-point.
We write the mass level $n_i$ as a subscript on the residue, so the four-point residue at mass level $n$ is written as $R_{n}$, the five-point residue at mass levels $n_1,n_2$ is written as $R_{n_1\,n_2}$, the six-point at levels $(n_1,n_2,n_3)$ as $R_{n_1\,n_2\,n_3}$, etc., according to the half-ladder topology in Fig.~\ref{fig:halfladder}.

Any candidate residues can describe a theory consistent with unitarity, with minimal degeneracy, only if there exists some real choice of the $\lambda$ couplings that reproduces them. In this case, Eq.~\eqref{eq:AKNu} leads to the four-point residues
\begin{equation}
\begin{aligned}\label{eq:string_residues}
    & R_0^\text{KN} = 1 \, , \\
    & R_1^\text{KN} = 1 - \alpha_0 - X_{2,4} \, , \\
    & R_2^\text{KN} = 1 - \frac{1}{2}(\alpha_0-2+X_{2,4})(\alpha_0-1+X_{2,4}) \, ,
\end{aligned}
\end{equation}
along with their higher-point counterparts given in the supplemental files \cite{supplementRstring}.

Performing this matching to the string residues in Eqs.~\eqref{eq:string_residues}, we find that $\alpha_0$ is fixed to $-1$.
This can be seen as follows.
Demanding $R_{2}^{\rm KN} = R_{2}^\lambda$ at four-point and $R_{2\,2}^{\rm KN} = R_{2\,2}^{\lambda}$ at five point allows us to fix 13 of the $\lambda$ couplings; these solutions are unique, up to trivial sign choices associated with $\mathbb{Z}_2$ flips of the exchanged states.
Then taking these solutions and plugging them into the six-point residues, we find that 
\begin{equation}
\lim_{X_{i,j}\rightarrow 0}\partial_{X_{2,4}}\partial_{X_{2,6}} (R_{2\,2\,2}^{\lambda} - R_{2\,2\,2}^{\rm KN}) = \frac{\alpha_0(1+\alpha_0)}{2(2+\alpha_0)}.
\end{equation}
Factorization requires that the right-hand side vanish, so we must have $\alpha_0 = 0$ or $-1$. The case of $\alpha_0 = -2$, where the $n=2$ state becomes massless, can be immediately ruled out by  the four-point residue at this level, where we find that factorization would require setting $\lambda_{1,0}^{0,0;0}$ to $\pm i/\sqrt{2}$, in violation of unitarity; this is consistent with Ref.~\cite{Arkani-Hamed:2023jwn}, which found that unitarity bounds $\alpha_0 \geq -1$ from the four-point partial wave expansion and from higher-point factorization at levels $\leq 1$.
Turning to the case of $\alpha_0 = 0$, we again find that factorization at four- and five-point on the all level-$2$ residues, $R_{2}^{\rm KN} = R_{2}^\lambda$  and $R_{2\,2}^{\rm KN} = R_{2\,2}^{\lambda}$, fixes 13 $\lambda$ couplings. When plugged into the six-point residue, we have
\begin{equation}
\begin{aligned}
&R^{\rm KN}_{2\,2\,2} - R^\lambda_{2\,2\,2} \\&= -\frac{1}{32}X_{2,4}X_{4,6}[D(2-X_{2,4})(2{-}X_{4,6}) \\&\qquad\qquad \qquad \;\; -8(5{+}X_{2,5}{-}X_{2,6}{-}X_{3,5}{+}X_{3,6})\\&\qquad \qquad \qquad \;\;{+}28X_{4,6}{+}2X_{2,4}(14{-}9X_{4,6}) ].
\end{aligned}
\end{equation}
This expression cannot vanish for all $X_{i,j}$ in any $D$, so $\alpha_0 = 0$ is also ruled out by factorization at minimal degeneracy.
Note, crucially, that this is not inconsistent with the existence of the superstring nor of tree-level $Z$-theory amplitudes.
In particular, the superstring is known to have a degenerate spectrum starting at level $n=2$ (see, e.g., Ref.~\cite{Basile:2024uxn}), and our results imply that $Z$-theory must have nontrivial degeneracy at $n=2$ as well.

These results leave $\alpha_0 = -1$, the Regge intercept of the bosonic string, the {\it unique choice} consistent with the Koba-Nielsen amplitude, unitarity, and minimal degeneracy through level $n=2$.
The consistent assignment of $\lambda$ couplings,\footnote{We computed these couplings by taking the $\alpha_0 \rightarrow-1$ limit of the case of arbitrary Regge intercept; if we instead reformulate the problem with massless propagators at level $n=1$, some of the couplings that are set to zero in the table can be taken to be free, as they decouple from the half-ladder massless amplitudes.} which is unique up to a sign flip for each exchanged state, is given in the following table.
\begin{center}
\bgroup
\def\arraystretch{1.0}
\begin{tabular}{c | c || l}\label{eq:family}
$(n_1, n_2)$ & $(s_1,s_2)_{j}$ & three-point couplings $\lambda^{s_1, s_2;j}_{n_1, n_2}$ \\ \hline \hline
$(0,0)$ & $(0,0)_0$ & 1 \\ \hline
$(0,1)$ & $(0,0)_0$ & 0 \\ \hline
$(0,1)$ & $(0,1)_0$ & $\sqrt{2}$ \\ \hline
$(0,2)$ & $(0,0)_0$ & $\sqrt{(26-D)/8(D-1)}$ \\ \hline
$(0,2)$ & $(0,1)_0$ & 0 \\ \hline
$(0,2)$ & $(0,2)_0$ & $\sqrt{2}$ \\ \hline
$(1,1)$ & $(0,0)_0$ & $1/2$ \\ \hline
$(1,1)$ & $(0,1)_0$ & 0 \\ \hline
$(1,1)$ & $(1,1)_{0,1}$ & 2, 1 \\ \hline
$(1,2)$ & $(0,0)_0$ & $0$ \\ \hline
$(1,2)$ & $(0,1)_0$ & $1/\sqrt{2}$ \\ \hline
$(1,2)$ & $(0,2)_0$ & $0$ \\ \hline
$(1,2)$ & $(1,0)_0$ & 0 \\ \hline
$(1,2)$ & $(1,1)_{0,1}$ & $0, 0$ \\ \hline
$(1,2)$ & $(1,2)_{0,1}$ & 2, 2 \\ \hline
$(2,2)$ & $(0,0)_{0}$ & $(9D-34)/8(D-1)$ \\ \hline
$(2,2)$ & $(0,1)_{0}$ & $0$ \\ \hline
$(2,2)$ & $(0,2)_{0}$ & $-\sqrt{(26-D)/4(D-1)}$ \\ \hline
$(2,2)$ & $(1,1)_{0,1}$ & $1$, $1/2$ \\ \hline
$(2,2)$ & $(1,2)_{0,1}$ & 0, $0$ \\ \hline
$(2,2)$ & $(2,2)_{0,1,2}$ & 2, 4, 1
\end{tabular}
\egroup
\end{center}

We conclude that through imposing factorization alone, among all $N$-point Koba-Nielsen amplitudes of arbitrary $\alpha_0$, and allowing for only minimal degeneracy, one straightforwardly recovers the precise Regge intercept of the bosonic string. This is not only suggestive of the rigidity of string theory, but also an indication that from the point of view of open string theory, the bosonic string is the only way of implementing minimal degeneracy up through level $2$.

\subsection{Deformations of the worldsheet integrand}\label{sec:level-two_Pu}

Let us now consider a deformation of the worldsheet, obtained by multiplying the integrand in the Koba-Nielsen form of the string amplitude by some form factor in terms of the $u$-variables, $P_N(u)$, in principle different for each multiplicity $N$, as shown in Eq.~\eqref{eq:AKNPu}.
Equivalently, such models can be thought of as satellite sums, with amplitudes giving by infinite sums over terms corresponding to string amplitudes at integer-shifted kinematics.
Various such $P(u)$ theories have been considered in the literature, notably in Refs.~\cite{Gross:1969db,Arkani-Hamed:2023jwn,Cheung_2025}.
In particular, Ref.~\cite{Arkani-Hamed:2023jwn} constructed $P(u)$ theories for the case of $\alpha_0 = 0$ in a manner such that factorization was guaranteed on the level-$0$ poles; factorization on the level-$1$ poles was found to constrain the ansatz.
Meanwhile, Ref.~\cite{Gross:1969db} built $P(u)$ theories in which, remarkably, factorization at all mass levels was guaranteed, albeit with imaginary couplings violating unitarity and significant degeneracy; such theories were found to be constrained by imposing unitarity of $N$-point scattering in Ref.~\cite{Cheung_2025}.

Here, we are interested in a different question to these past explorations: Requiring minimal degeneracy up through level $n=2$ as above, i.e., at most a single state for each spin $\leq n$ at mass level $n$, which $P(u)$ theories survive factorization constraints?
Taking a completely general $P_N(u)$ ansatz, that is, treating $P_N(u)$ as an arbitrary function of the $u$-variables at each multiplicity $N$ that is assumed to be at least twice-differentiable in each $u$-variable, and imposing factorization up through multiplicity $N=7$ for all levels $\leq 2$ with no degeneracy, we find that the only possibilities for a consistent solution are either the bosonic string itself, with $\alpha_0 = -1$---which is trivially a solution, with $P_N(u) = 1$, via the discussion in the previous section---or a {\it single} family of solutions\footnote{Specifically, imposing factorization through $N=7$, we find that the residues up through $N=6$ are uniquely fixed by the choice of couplings in the table. At seven point, the residues depend on the coupling $\lambda_{2,2}^{0,1;0}$, but we conjecture that imposing factorization on the $N>7$ amplitudes will fix this coupling as well. We leave further exploration of this remarkable family of residues at higher multiplicity to future work.} with couplings as in the following table, again modulo sign flips for each exchanged state.
\begin{center}
\bgroup
\def\arraystretch{1.0}
\begin{tabular}{c | c || l}\label{eq:family}
$(n_1, n_2)$ & $(s_1,s_2)_{j}$ & three-point couplings $\lambda^{s_1, s_2;j}_{n_1, n_2}$ \\ \hline \hline
$(0,0)$ & $(0,0)_0$ & 1 \\ \hline
$(0,1)$ & $(0,0)_0$ & 0 \\ \hline
$(0,1)$ & $(0,1)_0$ & $\sqrt{2}$ \\ \hline
$(0,2)$ & $(0,0)_0$ & 0 \\ \hline
$(0,2)$ & $(0,1)_0$ & 0 \\ \hline
$(0,2)$ & $(0,2)_0$ & $\sqrt{2}$ \\ \hline
$(1,1)$ & $(0,1)_0$ & 0 \\ \hline
$(1,1)$ & $(1,1)_{0,1}$ & 2, 1 \\ \hline
$(1,2)$ & $(1,0)_0$ & $-\frac{1}{2+\alpha_0}\sqrt{\frac{(1+\alpha_0)(6+5\alpha_0)}{2(2-3\alpha_0)}}$ \\ \hline
$(1,2)$ & $(1,1)_{0,1}$ & $0, -\sqrt{\frac{1+\alpha_0}{2(2+\alpha_0)}}$ \\ \hline
$(1,2)$ & $(0,2)_0$ & 0 \\ \hline
$(1,2)$ & $(1,2)_{0,1}$ & -2, -2 \\ \hline
$(2,2)$ & $(0,0)_{0}$ & $\frac{1}{2} + \frac{2(2-\alpha_0)(1+\alpha_0)}{(2+\alpha_0)^2(2-3\alpha_0)}$ \\ \hline
$(2,2)$ & $(0,2)_{0}$ & $\frac{1}{2+\alpha_0}\sqrt{\frac{2(1+\alpha_0)(6+5\alpha_0)}{2-3\alpha_0}}$ \\ \hline
$(2,2)$ & $(1,1)_{0}$ & $\frac{1}{2+\alpha_0}$ \\ \hline
$(2,2)$ & $(1,2)_{0,1}$ & 0, $\sqrt{\frac{2(1+\alpha_0)}{2+\alpha_0}}$ \\ \hline
$(2,2)$ & $(2,2)_{0,1,2}$ & 2, 4, 1
\end{tabular}
\egroup
\end{center}
Furthermore $\lambda_{2,2}^{1,1;1}$ and $\lambda_{2,2}^{0,1;0}$ are related according to
\begin{equation}
    \lambda_{2,2}^{1,1;1} + \sqrt{\frac{(2+\alpha_0)(6+5\alpha_0)}{2-3\alpha_0}}\lambda_{2,2}^{0,1;0} = \frac{4+\alpha_0}{2(2-3\alpha_0)} \, .
\end{equation}
Finally, in this family of solutions there is a constraint relating the dimension and $\alpha_0$:
\begin{equation}
D = 26-\frac{4(1+\alpha_0)(8+3\alpha_0)}{2+\alpha_0} \, .\label{eq:Dimconst}
\end{equation}
The reason that $D$ enters the factorization constraint is that the massive spin-two propagator at level $n=2$ depends on the spacetime dimension.
When $\alpha_0 = -1$, this family of solutions requires $D=26$, but this is an unnecessary constraint in this limit, since as we saw in the previous section the $\alpha_0 = -1$ case corresponding to the bosonic string in fact has only the constraint $D\leq 26$.
Interestingly, when $\alpha_0 = 0$, the constraint in Eq.~\eqref{eq:Dimconst} becomes $D=10$, which is suggestive of the superstring; however, in this case the residues in this family do not themselves match those of the superstring.
In fact, the residues in the family corresponding to the choices of couplings in the table can {\it only} match those of the Koba-Nielsen family described by Eq.~\eqref{eq:AKNu} precisely when $\alpha_0 = -1$.
The new family of residues described here could be thought of as an $\alpha_0$-dependent deformation of the bosonic string, distinct from the Koba-Nielsen residues, and one that uniquely permits no degeneracy among states of equal spin and mass for levels $n\leq 2$. 
However, the requirement of integer dimension means that Eq.~\eqref{eq:Dimconst} reduces this deformation to a discrete family.
Our condition $\alpha_0 \geq -1$ forbidding spinning tachyons means that $D \leq 26$ in this family, and reality of the $\lambda$ couplings as required by unitarity restricts us to the window $-1\leq \alpha_0 < 2/3$.

Imposing factorization implies various constraints on $P_N(u)$ and its derivatives.
To identify the unique family described above, with no degeneracy through level $n=2$, required going to multiplicity $N=7$, in which case factorization constrains up to second derivatives in $P_{4,5,6,7}(u)$.
These constraints can be read off by comparing the ansatz residues to those in the $P(u)$ models.
In the supplemental files \cite{supplementRP}, we have given the residues through seven-point and through mass level $n=2$ for theories with arbitrary worldsheet deformation $P_N(u)$, which can be compared to the ansatz residues, also included in the supplemental files \cite{supplementRansatz}.
We have also given in the supplemental files \cite{supplementRspecial} the residues for the unique family identified above through six-point and mass level $n=2$.
We leave to future work the compelling question of whether these residues correspond to a fully factorizable model, consistent with unitarity, that can be extended to arbitrarily high multiplicity and mass level.

\section{Massless scattering}\label{sec:massless_scattering}

Massless external states are the relevant setting for (non-tachyonic) open superstrings. Ultimately, we would like to extend our methods to study graviton scattering described by deformations of the Virasoro-Shapiro amplitude (and their higher-point counterparts), but for the time being we focus on deformations in the open sector. Besides their physical relevance, massless external states bring simplified kinematics, which however comes at a price: assuming non-degeneracy at level two would exclude superstrings from our theory space. Taking level-two degeneracy of the spectrum into account and working through the exercise of explicit factorization is possible, but it would significantly complicate the analysis beyond the scope of this paper. Hence, in this section we focus on a subset of bounds that are {\it implied by} consistent factorization, but which remarkably do {\it not} require any assumptions about spectral degeneracies. These are the multipositivity bounds introduced in Ref.~\cite{Cheung_2025}. We will consider the simplest version of these bounds, in which we take a multi-soft limit where the incoming momenta $p_3,p_4,\ldots,p_{N-2}$ along the spine of the half-ladder vanish (hence requiring massless external states).\footnote{Generalizations of the multi-soft bounds include Mellin-transformed external states and Hermitian reweightings of the internal three-point couplings of the half-ladder diagram, explored at length in Ref.~\cite{Cheung_2025}, as well as conjectured constructions involving certain more general topologies. We leave the question of a saddle-point large-$n$ treatment of these more general bounds to future work.} Effectively, this choice reduces the problem to a two-to-two scattering in a soft background for any number of external particles. As shown in Ref.~\cite{Cheung_2025}, this approach has the virtue of allowing us to impose multipositivity bounds for the whole infinite tower of mass levels and arbitrary numbers of external particles, producing genuinely novel positivity bounds that cannot arise from any finite number of states. \\

The $N$-point kinematics for $N > 4$ in the soft limit $p_3, \dots p_{N-2} \to 0$ is determined by two Mandelstam variables: the center-of-mass energy $s = -2p_1\cdot p_2$ along the spine, and the overall momentum transfer $t = -2p_2 \cdot p_{N-1} =-X_{2,N}$. All external states are massless, $p_i^2=0$ for $i=1,2,\ldots,N-1,N$. In this factorization channel, all resonances appear at the same level $n$, and the residues, which we write as $R_{N,n}(t)$, depend only on $t$. For these specifications, Ref.~\cite{Cheung_2025} found that the residues $R_{N,n}(t)$ can be written as expectation values of powers of Hermitian operators on the single-particle state space, which encode the effect of three-point vertices in the soft limit. Specifically, for $t\geq 0$,
\begin{equation}
    \begin{aligned}
        R_{4,n} &= \braket{n|n} \\
        R_{5,n} &= \bra{n} \mathcal{O}_n \ket{n} \\ 
        R_{6,n} &= \bra{n} \mathcal{O}_n^2 \ket{n}
    \end{aligned}
\end{equation}
and so on. The definitions of ${\cal O}_n$ and $\ket{n}$ can be found in Ref.~\cite{Cheung_2025}; for our purposes, it will be sufficient to note that taking $\ket{n}$ to be a well-defined vector and ${\cal O}_n$ to be Hermitian are precisely implied by the reality of the $\lambda$ three-point couplings along the spine of the half-ladder diagram.
As the inner product should be consistent, we obtain the conditions $\braket{n|n} \geq 0$ and $\bra{n} \mathcal{O}_n^2 \ket{n} \geq 0$. Furthermore, the variance dictates that $\bra{n} \mathcal{O}_n^2 \ket{n}  \braket{n|n}- (\bra{n} \mathcal{O}_n \ket{n})^2 \geq 0 $ , which means that $R_{6,n}R_{4,n}-R_{5,n}^2 \geq 0$. 
As shown in Ref.~\cite{Cheung_2025}, these bounds generalize to an arbitrarily high number of legs $N$, summarized in the Hankel matrix comprising powers of ${\cal O}_n$,
\begin{equation}
\begin{pmatrix}
    R_{4,n} & R_{5,n} & R_{6,n} &  \cdots \\
    R_{5,n} & R_{6,n} & R_{7,n} & \\
    R_{6,n} & R_{7,n} & R_{8,n} &  \\
      \vdots &   &    & \ddots  \\
\end{pmatrix},\label{eq:Hankel}
\end{equation}
where unitarity requires that all principal minors of Eq.~\eqref{eq:Hankel} be positive semidefinite for $t\geq 0$. 

\subsection{Multipositivity at large level}\label{sec:large-level}

Soft kinematics allow us to explore factorization of superstring deformations with massless external states and arbitrary degeneracy to all levels. In order to compute the soft residues at large level $n \gg 1$, we observe that the kinematic configuration dramatically simplifies the $N$-point amplitude in Eq.~\eqref{eq:AKNPu} to
\begin{eqaed}
    A_N = \int_0^{\infty} \prod_{i=3}^{N-1}\frac{dy_{1,i}}{y_{1,i}} \, u_{2,N}^{-t}(1-u_{2,N})^{-s} \, P_N(u) \, .
\end{eqaed}
This expression looks very similar to the familiar Veneziano amplitude, and highlights the intuitive characterization of this soft limit as a four-point amplitude in a soft background \cite{Cheung_2025}. The residue at level $s=n$ can then be computed as the residue of the integrand at $y_{1,i}=0$, as we have done thus far. Clearly, consistent factorization up to any given level can be achieved by choosing the deformations $P_N(u)$ such that all the residues up to the chosen level are precisely the undeformed string residues, hiding the deformations up at yet higher, untested mass levels. It is thus desirable to derive factorization constraints to arbitrarily high levels, leveraging the rigidity of the higher-spin tower of resonances. Owing to their elegant structure, multipositivity bounds allow us to explore arbitrarily high-point and high-level factorization, albeit not in complete kinematic generality; with this trade-off in mind, we can thus make progress by studying $R_{N,n}(t)$ for mass level $n \gg 1$, for which we now derive an asymptotic expression for any $P_N(u)$. In detail, we consider the limit $n \gg 1$ with $z \equiv t/n$ fixed. This is analogous to the hard scattering limit studied in Refs.~\cite{Gross:1987kza, Gross:1987ar}.

If $P_N(u)=1$, the residues are of course simply those of the string, which independent of the number of external particles $N$ remarkably satisfy~\cite{Cheung_2025}
\begin{equation}
R_{N,n}^{\text{KN}}(t) = \binom{n+t}{n}.\label{eq:Rstringbin}
\end{equation}
This result follows from the expression \cite{Arkani-Hamed:2023lbd}
\begin{eqaed}\label{eq:u2N}
    u_{2,N} = \frac{1+\sum_{k=0}^{N-5}\prod_{i=0}^k y_{1,N-1-i}}{1+\sum_{k=0}^{N-4}\prod_{i=0}^k y_{1,N-1-i}} ,
\end{eqaed}
which forces the residue of the integrand (evaluated at level $s=n$) at $y_{1,i}=0$ to be given by the $O(y_{1,i}^0)$ part of the rest of the integrand, after peeling off the $dy/y$ measure, which in turn by Eq.~\eqref{eq:u2N} is given by the coefficient of $\prod_{i=3}^{N-1} y_{1,i}^n$ in the combination $(1+\dots+\prod_{i=3}^{N-1} y_{1,i})^{n+t}$, where we highlighted the only term in the integrand containing $y_{1,3}$. For general $P_N(u)$, the saddle-point computation remains analogous: for $z=t/n>0$ fixed, the Koba-Nielsen factor is the only part of the integrand that scales in the large-$n$ limit. Therefore the leading-order result for the residue is given by the string residue $R_{N,n}^{\text{KN}}(t)$ multiplied by $P_N(u)$ evaluated at the saddle point, since the Gaussian fluctuations around the saddle are already included in $R_{N,n}^{\text{KN}}(t)$. The relevant integral that computes the residue is
\begin{eqaed}
\hspace{-2mm}    R_{N,n}(nz) {=} \oint_0 \prod_{k=3}^{N-1} \frac{dy_{1,k}}{2\pi i y_{1,k}} \, u_{2,N}^{-nz}(1{-}u_{2,N})^{-n} P_N(u),\hspace{-2mm}
\end{eqaed}
taking each contour to be a small circle around the origin. The part of the integrand that scales in the large-$n$ limit is $\exp(n S(u,z))$, where
\begin{eqaed}
    e^{-S(u,z)} = u_{2,N}^z (1-u_{2,N}) \, .
\end{eqaed}
The saddle-point equations then read
\begin{eqaed}
    \left(\frac{z}{u_{2,N}}-\frac{1}{1-u_{2,N}}\right) \frac{\partial u_{2,N}}{\partial y_{1,i}} = 0 .\label{eq:saddle}
\end{eqaed}
The gradient term $\partial u_{2,N}/\partial y_{1,i}$ can vanish only in the limit $y_{1,3}\rightarrow\infty$, where $u_{2,N}\rightarrow 0$ and the integrand vanishes rapidly; such a limit cannot be the saddle, as this would yield a vanishing residue.
Thus, we conclude that the saddle-point surface is defined by the vanishing of the factor in parentheses in Eq.~\eqref{eq:saddle}, which happens when $u_{2,N}=z/(1+z)$. In particular, for $N=4$ the large-$n$ residue is $R_{4,n}(nz) \sim R_{4,n}^{\text{KN}}(nz) \, P_4^*$, where hereafter we write $P_4^* \equiv P_4\big(u_{1,3} {=} \frac{1}{1+z},u_{2,4} {=} \frac{z}{1+z}\big)$. Positivity of $R_{4,n}(n z)$, required by unitarity, implies $P_4 \geq 0$ in its domain, since $z$ is arbitrary. In fact, as we now show, it turns out that this is the only positivity bound arising from the leading-order analysis. For $N > 4$ the equation $u_{2,N} = \frac{z}{1+z}$ can be solved for $y_{1,3}$, which leads to
\begin{eqaed}\label{eq:y13_saddle}
    y_{1,3} = y_{1,3}^{\text{saddle}} = \frac{1+\sum_{k=0}^{N-5} \prod_{i=0}^k y_{1,N-1-i}}{z \prod_{i=4}^{N-1}y_{1,i}} \, .
\end{eqaed}
Therefore, we can parameterize a neighborhood of the saddle-point hypersurface by the other $y_{1,i}$ and the deviation $y_{1,3} = y_{1,3}^{\text{saddle}} + \delta y_{1,3}$. To leading order in the saddle-point approximation, one has $S(u,z) \sim S(u^\text{saddle},z) + \frac{(1+z)^3}{2z} \, (\partial_{y_{1,3}}u_{2,N})^2 \, \delta y_{1,3}^2$, where the derivative on the saddle-point hypersurface simplifies to
\begin{eqaed}
\begin{aligned}
    \frac{\partial u_{2,N}}{\partial y_{1,3}} \bigg|_{\text{saddle}} &= - \, \frac{u_{2,N}(1-u_{2,N})}{y_{1,3}} \bigg|_{\text{saddle}} \\&= - \, \frac{z}{(1+z)^2 y_{1,3}^\text{saddle}} .
    \end{aligned}
\end{eqaed}
The complex Gaussian integral over $\delta y_{1,3}$ is then controlled by the direction of steepest descent, and the only source of complex phases is $y_{1,3}^\text{saddle}$. Rotating $\delta y_{1,3}$ along the direction of steepest descent,  the Gaussian exponent becomes $-\frac{n}{2}\frac{z}{1+z} \big|\delta y_{1,3}/y_{1,3}^\text{saddle}\big|^2$. The Gaussian integral over $\delta y_{1,3}$ then leaves the saddle-point expression for the residue,
\begin{eqaed}
    \sqrt{\frac{1+z}{2\pi nz}} \oint_0 \prod_{i=4}^{N-1} \frac{dy_{1k}}{2\pi i y_{1k}} \, P_N(u^\text{saddle}) \, .
\end{eqaed}
Since the $u$-variables are holomorphic in a neighbourhood of the origin, the integral simply evaluates to $P_N(u^*)$ where $u^* = u^{\text{saddle}}(y_{1,i}=0)$. But this corresponds to a massless complete factorization channel down to four-point.
Provided that we have chosen a deformation $P_N(u)$ that is automatically consistent with massless factorization, as in Refs.~\cite{Arkani-Hamed:2023jwn,Gross:1969db}, we have $P_N(u^*) = P_4^*$. Hence, for $n \gg 1$,
\begin{eqaed}
    R_{N,n}(nz) \sim \sqrt{\frac{1+z}{2\pi n z}} \, e^{n S(u^*,z)} \, P_4^*.
\end{eqaed}
Reinstating the saddle-point exponential prefactor $\exp\left(n S(u^*,z)\right) = (1+z)^{n(1+z)}z^{-nz}$ and comparing with Eq.~\eqref{eq:Rstringbin} in the large-$n$ limit using the Stirling approximation, we find 
\begin{equation}
 R_{N,n}(nz) \sim R_{N,n}^\text{KN}(nz) \, P_4^*
\end{equation}
as anticipated.

The upshot of this calculation is the positivity bound $P_4 \geq 0$. The matrix of soft residues has equal entries to leading order in the large-level limit, and thus it does not yield additional bounds. However, one expects subleading corrections to probe higher-point processes close to massless factorization channels, whose contributions may control the (vanishing at leading order) minors of the Hankel matrix of multipositivity bounds in Eq.~\eqref{eq:Hankel}.

\subsection{First subleading correction}\label{sec:first_subleading_correction}

Computing subleading corrections to the above saddle-point calculation is in principle straightforward, albeit tedious. One needs to retain higher-order terms in $\delta y_{1,3}$ both in the exponential and the rest of the integrand. Then, rescaling $\delta y_{1,3} \to \epsilon$ in order to recast the quadratic part of the exponential into its canonical form $\frac{1}{2} \epsilon^2$, the steepest-descent direction is selected by rotating $\epsilon \to i \epsilon$. Since the exponent was proportional to $n$ in the original variables, after rescaling each higher-order term corrects it by powers of $\epsilon/\sqrt{n}$, and analogously for the rest of the integrand. Thus, since the Gaussian distribution has vanishing odd moments, the first subleading correction enters at $O(1/n)$ and arises from the contributions up to $O(\epsilon^4)$ from the Taylor expansion of the exponential, and from the $O(\epsilon^2)$ contribution in the rest of integrand.\footnote{The difference in powers of $\epsilon$ is ascribed to the fact that the exponentiated terms carry an extra factor of $\epsilon^2$. However, the scaling with $n$ of these subleading terms in the final result is the same.}
 All in all, the result takes the form
\begin{eqaed}\label{eq:subleading_corr}
    R_{N,n}(nz) \sim \sqrt{\frac{1+z}{2\pi n z}} \, e^{n S(u^*,z)} \left(P_4^* + \frac{1}{n} \, R_N^{(1)}(z)\right) ,
\end{eqaed}
where the correction 
\begin{eqaed}
 \hspace{-1mm}   R_N^{(1)}(z) = - \frac{1+z+z^2}{12z(1+z)} \, P_4^* - \frac{2z P_N^{(1)} + (1+z) P_N^{(2)}}{2z^3}
\end{eqaed}
is written in terms of the limits
\begin{eqaed}\label{eq:composite_derivatives}
    P_N^{(k)} \equiv \lim_{y_{1,i \to 0}} \left(\prod_{i=4}^{N-1}\frac{1}{y_{1,i}^k}\right) \frac{\partial^k P_N}{\partial y_{1,3}^k}\bigg|_{y_{1,3}=y_{1,3}^\text{saddle}} .
\end{eqaed}
Above, we have required as before that the  $P_N$ have been chosen such that massless factorization down to lower-point amplitudes is mechanically satisfied. In particular, taking $P_4=1$, we recover the first subleading correction the large-$n$ expansion of $R_{N,n}^\text{KN}(nz)$, which one can check using Eq.~\eqref{eq:Rstringbin} and corrections to Stirling's approximation. Hence, at this order we can recast Eq.~\eqref{eq:subleading_corr} into the simpler form
\begin{eqaed}
    \frac{R_{N,n}(nz)}{R_{N,n}^\text{KN}(nz)} \sim P_4^* - \frac{2z P_N^{(1)} + (1+z) P_N^{(2)}}{2z^3n} \, .
\end{eqaed}
The complexity is hidden in the composite derivatives and limits in Eq.~\eqref{eq:composite_derivatives}, while the remaining structure is universal. The $2\times 2$ principal minors for contiguous blocks yield a new family of positivity bounds for all $z$, as anticipated,\footnote{More precisely, the whole left-hand side should be multiplied by $P_4^*$. Assuming that $P_4$ never vanishes yields Eq.~\eqref{eq:subleading_2x2_bound}.}
\begin{eqaed}\label{eq:subleading_2x2_bound}
    & (1+z)(2P^{(2)}_{N+1}-P^{(2)}_{N}-P^{(2)}_{N+2}) \\
    & + 2z (2P^{(1)}_{N+1}-P^{(1)}_{N}-P^{(1)}_{N+2}) \geq 0 .
\end{eqaed}

\subsection{Satellite theories}\label{sec:satellite} 

As a testing ground for our factorization constraints, let us consider the simplest of the satellite $P(u)$-theories defined in Ref.~\cite{Gross:1969db}. This is the so-called $\tilde{f}_4$-theory, since the tree-level S-matrix for massless external particles is completely determined by a single analytic function $\tilde{f}_4(x)$ of a single variable on the interval $0 \leq x \leq 1$. This function determines the $P_N(u)$ integrands according to
\begin{eqaed}\label{eq:tilde-f_4_theory}
    P_N(u) = \exp \left[\sum_{k=1}^N u_{k,k+2}(1{-}u_{k,k+2})\tilde{f}_4(u_{k,k+2})\right] .
\end{eqaed}
More complicated deformations involve additional functions $\tilde{f}_N(x)$ for $N>4$, and the corresponding structure of Eq.~\eqref{eq:tilde-f_4_theory} guarantees factorization to arbitrary $N$-point (albeit possibly with imaginary couplings).

We now apply the bounds derived in Sec.~\ref{sec:first_subleading_correction} to this theory. Remarkably, we will be able to show that it {\it cannot} deform the four-point amplitude. Since each $P_N$ is an exponential, and thus by definition nonnegative, the leading large-level bound $P_4 \geq 0$ is automatically satisfied. This makes the $\tilde{f}_4$-theory in particular a good candidate to test the subleading bound in Eq.~\eqref{eq:subleading_2x2_bound}. Another reason why multipositivity bounds are an effective tool to probe the consistency of these deformations is that they ignore degeneracy, which can be very large in such models~\cite{Gross:1969db}. \\

The $\tilde{f}_4(x)$ deformation was studied, in the case of massless external states, using multipositivity bounds through mass level $n=2$ in Ref.~\cite{Cheung_2025}, which yield $\tilde{f}_4(0) = \tilde{f}_4(1) = \tilde{f}_4'(0) = \tilde{f}_4'(1) = 0$. 
Here, we will instead use our large-level analysis above to place much more stringent constraints on this theory.
Let us define
\begin{equation}
\tilde{f}_4(x) = \tilde f_+(x) + \tilde f_-(x),
\end{equation}
where we have isolated the (anti-)symmetric parts under $x\rightarrow 1-x$, so that $\tilde f_\pm (1-x) = \pm \tilde f_\pm (x)$.
Without loss of generality, let us define 
\begin{equation}
\tilde f_+(x) = \frac{\log(\sqrt{2+w(x)}/2)}{x(1-x)}
\end{equation}
for some function $w(x)=w(1-x)$.
We find that, for the $\tilde f_4$-theory, $\tilde f_-$ drops out of $P_4(x,1-x) = (2+w(x))^2/16$ (where here we use the four-point $u$ equation $u_{13} + u_{24} = 1$ to write $P_4(u_{13},u_{24})$ as simply $P_4(x,1-x)$).
Imposing the normalization in Ref.~\cite{Gross:1969db}, $P_4(1,0)=P_4(0,1)=1$, we have $w(0)=w(1)=2$.
The multipositivity bound relating four-, five-, and six-point residues in Eq.~\eqref{eq:subleading_2x2_bound}, which we derived from the first subleading corrections around the saddle point at large mass level, dramatically simplifies in these variables to
\begin{eqaed}
    w(x) w''(x) + (w'(x))^2 \leq 0 .\label{eq:wbound}
\end{eqaed}
But
\begin{equation}
w(x) w''(x) + (w'(x))^2 = \frac{d}{dx}(w(x) w'(x)),
\end{equation}
and since $w'(0)=4(\tilde f_4 (0) + \tilde f_4(1)) = 0$ by the bounds in Ref.~\cite{Cheung_2025}, we have the identity
\begin{equation}
\hspace{-1mm}\!\int_0^1 \!\frac{d}{dx} (w(x)w'(x)) \, dx = w(1)w'(1){-}w(0)w'(0) \,{=}\, 0.\label{eq:int}
\end{equation}
Since the integrand is of definite sign by the constraint in Eq.~\eqref{eq:wbound}, it follows that the bound is exactly saturated,
\begin{eqaed}
    w(x) w''(x) + (w'(x))^2 = 0 ,
\end{eqaed}
and therefore $w(x) = \sqrt{c_0+c_1x}$ for some constants $c_{0,1}$. Using that $w(x) = w(1-x)$ and $w(0) = 2$, we conclude that $w(x) = 2$ identically, and thus $\tilde f_+(x) = 0$. This conclusion implies that unitarity forbids the massless $\tilde f_4$-theory from inducing {\it any} four-point deformation of the string amplitude, and moreover heavily constrains higher-point deformations by removing the even contribution to $\tilde{f}_4$. It is conceivable that an analysis along the same lines for higher $N > 4$, possibly accompanied by additional next-to-next-to-leading order bounds at large mass level, would constrain the odd part $\tilde f_-(x)$. 
We leave further constraints on yet more general string deformations in Ref.~\cite{Gross:1969db} to future work.

\section{Discussion}\label{sec:conclusions}

The results we have presented in this paper showcase the constraining power of unitarity of the S-matrix in the form of multiparticle factorization. In particular, fixing the Regge intercept for the bosonic string and ruling out the simplest satellite deformation for superstrings are further pieces of evidence to the effect that string theory may be the unique weakly coupled tree-level scattering amplitude obeying certain UV properties, thanks to its rigid and rich higher-spin structure. Unlike gluon scattering, it is famously not possible to UV-complete gravity within field theory.
Conversely, gravity is more highly constrained, where (unlike the case in field theory) it can be shown that any weakly coupled completion {\it must} occur at tree level~\cite{Cheung:2016wjt}. Therefore, though we did not directly examine closed-string deformations, seeking a weakly coupled completion of gravity suggests the usefulness of seeking enhanced UV behavior through tree-level mechanisms in the gauge sector as well, as indeed occurs in open string amplitudes. In this arena, the relevant multiparticle amplitudes in a consistent tree-level S-matrix are meromorphic functions with simple poles and polynomial residues, all related to each other by consistent factorization. This mathematical reflection of the rigidity of physical principles provides concrete and effective tools to advance our understanding of string theory and perhaps motivate its inevitability. To this end, it would be very interesting to further develop our methods into a systematic framework to exploit the higher-spin tower and the large-level limit simultaneously, enabling yet more powerful constraints to be derived. Another clear direction for further research is the extension of our methods to deformations of closed-string amplitudes. In this case the cross-ratio $u$-variables appear in a more complicated fashion~\cite{Brown:2019wna}, and their combinatorial role in computing residues on factorization channels and classifying the possible deformations at the integrand level remains elusive. It seems plausible that adapting the approach of Ref.~\cite{Arkani-Hamed:2023lbd} to the geometry of closed (punctured) Riemann surfaces and the combinatorics of full crossing symmetry could lead to more insights. Finally, it would be interesting to compare our results for flat-spacetime amplitudes to the growing literature on anti-de Sitter string amplitudes~\cite{Alday:2022xwz, Alday:2023jdk, Alday:2023mvu, Alday:2024ksp, Alday:2024yax, Wang:2025owf}. Specifically, the kinematic dependence of the deformations is characteristic of the presence of a curved background, which prompts the search for a clear-cut methodology to separate the background from the dynamical deformations to be constrained by the bootstrap.

\bigskip

\section*{Acknowledgments}
\noindent We thank Nima Arkani-Hamed, Cliff Cheung, Cl\'{e}ment Dupont, Carolina Figueiredo, Waltraut Knop, Chrysoula Markou, Giulio Salvatori, and Stephan Stieberger for useful discussions. 
The work of I.B. is supported by the Origins Excellence Cluster and the German-Israel-Project (DIP) on Holography and the Swampland. 
G.N.R. is supported by the James Arthur Postdoctoral Fellowship at New York University.

\bibliographystyle{utphys-modified}
\bibliography{HigherPoint}

\end{document}